\documentclass{article}

\usepackage{authblk}
\usepackage[utf8]{inputenc}
\usepackage{main} 
\usepackage{microtype}
\usepackage{times}
\usepackage{latexsym}

\usepackage{amsmath}
\usepackage{amssymb}
\usepackage{amsfonts}
\usepackage{mathtools}
\usepackage{bm}

\usepackage{graphicx}
\usepackage{subcaption}
\usepackage{float}
\usepackage{wrapfig}

\usepackage{booktabs}
\usepackage{array}
\usepackage{multirow}
\usepackage{multicol}
\usepackage{makecell}
\usepackage{longtable}
\usepackage{tabularx}
\usepackage{colortbl}
\usepackage{arydshln}

\usepackage{color}
\usepackage{xcolor}
\definecolor{mydarkblue}{rgb}{0,0.08,0.45}
\definecolor{wkblue}{rgb}{0.2, 0.3, 0.6}
\definecolor{meta-color}{rgb}{0.5, 0.5, 0.5}
\definecolor{bgblue}{RGB}{245,243,253}
\definecolor{ttblue}{RGB}{91,194,224}
\definecolor{mybrown}{RGB}{128,64,0}
\definecolor{titlecolor}{HTML}{4c9cff}
\definecolor{coolblue3}{rgb}{0.91, 0.94, 0.98}
\definecolor{myblue}{rgb}{0.9, 0.1, 0.94}
\definecolor{mygreen}{rgb}{0.64, 0.56, 0.88}
\definecolor{myyellow}{rgb}{0.68, 0.6, 0.1}
\definecolor{fancygreen}{rgb}{0.33, 0.68, 0.20}
\definecolor{salmon}{rgb}{0.94, 0.52, 0.49}
\definecolor{tablegreen}{rgb}{0.82, 0.94, 0.75}
\definecolor{tableblue}{rgb}{0.81, 0.90, 0.94}
\definecolor{tablered}{rgb}{0.97, 0.85, 0.85}
\definecolor{tableorange}{rgb}{0.96, 0.85, 0.81}

\usepackage{enumitem}
\usepackage{footnote}
\usepackage{lipsum}
\usepackage{setspace}

\usepackage{geometry}
\usepackage{fancyhdr}
\usepackage{lscape}
\usepackage{rotating}

\usepackage{algorithm}
\usepackage{algorithmic}

\usepackage[most]{tcolorbox}
\usepackage[framemethod=tikz]{mdframed}
\usepackage{awesomebox}

\usepackage{natbib}
\usepackage{url}
\usepackage[colorlinks=true,linkcolor=mydarkblue,citecolor=mydarkblue,filecolor=mydarkblue,urlcolor=mydarkblue]{hyperref}

\usepackage{bbding}
\usepackage{imakeidx}
\makeindex
\usepackage[toc]{multitoc}
\usepackage[edges]{forest}
\usepackage[normalem]{ulem}
\usepackage{fontawesome5}
\usepackage{blindtext}
\usepackage{pgfplots}
\usepackage{pgfplotstable}

\usepackage{listings}
\usepackage{listingsutf8}
\newcommand\JSONnumbervaluestyle{\color{blue}}
\newcommand\JSONstringvaluestyle{\color{red}}

\newif\ifcolonfoundonthisline

\makeatletter
\lstdefinestyle{json}
{
  showstringspaces    = false,
  keywords            = {false,true},
  alsoletter          = 0123456789.,
  morestring          = [s]{"}{"},
  stringstyle         = \ifcolonfoundonthisline\JSONstringvaluestyle\fi,
  MoreSelectCharTable =%
    \lst@DefSaveDef{`:}\colon@json{\processColon@json},
  basicstyle          = \ttfamily,
  keywordstyle        = \ttfamily\bfseries,
}

\newcommand\processColon@json{%
  \colon@json%
  \ifnum\lst@mode=\lst@Pmode%
    \global\colonfoundonthislinetrue%
  \fi
}

\lst@AddToHook{Output}{%
  \ifcolonfoundonthisline%
    \ifnum\lst@mode=\lst@Pmode%
      \def\lst@thestyle{\JSONnumbervaluestyle}%
    \fi
  \fi
  \lsthk@DetectKeywords%
}

\lst@AddToHook{EOL}%
  {\global\colonfoundonthislinefalse}
\makeatother

\newtcolorbox{myboxi}[1][]{
  breakable,
  title=#1,
  colback=red!5,
  colbacktitle=red!5,
  coltitle=black,
  fonttitle=\bfseries,
  bottomrule=0pt,
  toprule=0pt,
  leftrule=2pt,
  rightrule=2pt,
  titlerule=0pt,
  arc=0pt,
  outer arc=0pt,
  colframe=red,
}

\newtcolorbox{myboxnote}[1][]{
  breakable,
  title=#1,
  colback=orange!0,
  colbacktitle=orange!0,
  coltitle=black,
  fonttitle=\bfseries,
  bottomrule=0pt,
  toprule=0pt,
  leftrule=2pt,
  rightrule=2pt,
  titlerule=0pt,
  arc=0pt,
  outer arc=0pt,
  colframe=orange,
}

\newtcolorbox{myboxii}[1][]{
  breakable,
  freelance,
  title=#1,
  colback=white,
  colbacktitle=white,
  coltitle=black,
  fonttitle=\bfseries,
  bottomrule=0pt,
  boxrule=0pt,
  colframe=white,
  overlay unbroken and first={
  \draw[red!75!black,line width=3pt]
    ([xshift=5pt]frame.north west) --
    (frame.north west) --
    (frame.south west);
  \draw[red!75!black,line width=3pt]
    ([xshift=-5pt]frame.north east) --
    (frame.north east) --
    (frame.south east);
  },
  overlay unbroken app={
  \draw[red!75!black,line width=3pt,line cap=rect]
    (frame.south west) --
    ([xshift=5pt]frame.south west);
  \draw[red!75!black,line width=3pt,line cap=rect]
    (frame.south east) --
    ([xshift=-5pt]frame.south east);
  },
  overlay middle and last={
  \draw[red!75!black,line width=3pt]
    (frame.north west) --
    (frame.south west);
  \draw[red!75!black,line width=3pt]
    (frame.north east) --
    (frame.south east);
  },
  overlay last app={
  \draw[red!75!black,line width=3pt,line cap=rect]
    (frame.south west) --
    ([xshift=5pt]frame.south west);
  \draw[red!75!black,line width=3pt,line cap=rect]
    (frame.south east) --
    ([xshift=-5pt]frame.south east);
  },
}

\mdfdefinestyle{mystyle}{%
  rightline=true,
  innerleftmargin=10,
  innerrightmargin=10,
  outerlinewidth=3pt,
  topline=false,
  rightline=true,
  bottomline=false,
  skipabove=\topsep,
  skipbelow=\topsep
}

\usepackage{pifont}

\DeclareCaptionFont{black}{\color{black}}

\newenvironment{itemize*}%
 {\leftmargini=10pt\begin{itemize}%
  \setlength{\itemsep}{0pt}%
  \setlength{\parskip}{0pt}%
  }%
 {\end{itemize}}
\newenvironment{enumerate*}%
 {\begin{enumerate}%
  \setlength{\itemsep}{0pt}%
  \setlength{\parskip}{0pt}}%
 {\end{enumerate}}

\usepackage{etoolbox}
\newcounter{bibcount}
\makeatletter
\patchcmd{\@lbibitem}{\item[}{\item[\hfil\stepcounter{bibcount}{[\thebibcount]}}{}{}
\renewcommand\NAT@bibsetup%
  [1]{\setlength{\leftmargin}{\bibhang}\setlength{\itemindent}{-\parindent}%
      \setlength{\itemsep}{\bibsep}\setlength{\parsep}{\z@}}
\makeatother

\definecolor{myblue}{rgb}{0.9, 0.1, 0.94}
\definecolor{mygreen}{rgb}{0.64, 0.56, 0.88}
\definecolor{myyellow}{rgb}{0.68, 0.6, 0.1}
\definecolor{fancygreen}{rgb}{0.33, 0.68, 0.20}
\definecolor{salmon}{rgb}{0.94, 0.52, 0.49}
\definecolor{tablegreen}{rgb}{0.82, 0.94, 0.75}
\definecolor{tableblue}{rgb}{0.81, 0.90, 0.94}
\definecolor{tablered}{rgb}{0.97, 0.85, 0.85}
\definecolor{tableorange}{rgb}{0.96, 0.85, 0.81}

\begin{document}

\title{daVinci-Dev: Agent-native Mid-training for Software Engineering}

\author[2,3]{Ji Zeng}
\author[1,3]{Dayuan Fu}
\author[1,3]{Tiantian Mi}
\author[2,3]{Yumin Zhuang}
\author[3]{Yaxing Huang}
\author[1,2,3]{Xuefeng Li}
\author[3]{Lyumanshan Ye}
\author[1,3]{Muhang Xie}
\author[2,3]{Qishuo Hua}
\author[1,3]{Zhen Huang}
\author[1,2,3]{Mohan Jiang}
\author[1,3]{Hanning Wang}
\author[2,3]{Jifan Lin}
\author[3]{Yang Xiao}
\author[1,3]{Jie Sun}
\author[2,3]{Yunze Wu}
\author[1,2,3]{Pengfei Liu\textsuperscript{†}}
\affil{SII 
\quad \textsuperscript{2}SJTU 
\quad \textsuperscript{3}GAIR
}

\maketitle

\pagestyle{fancy}
\thispagestyle{fancy}
\fancyhead{}
\lhead{
  \raisebox{-0.3cm}{\includegraphics[height=0.95cm]{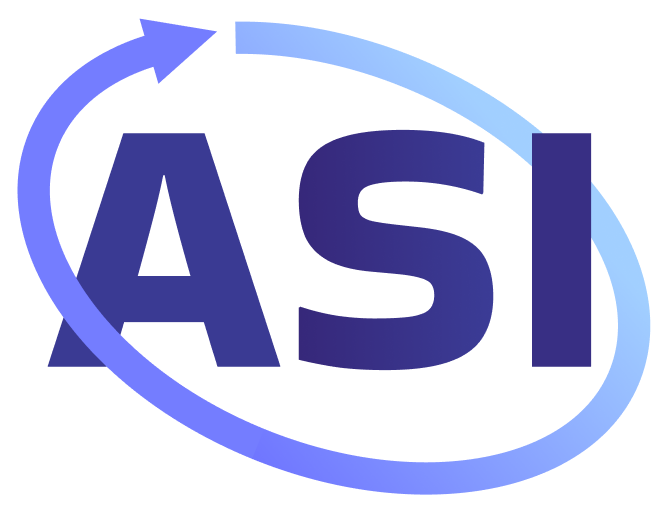}}
}
\rhead{%
  \raisebox{-0.2cm}{\includegraphics[height=0.7cm]{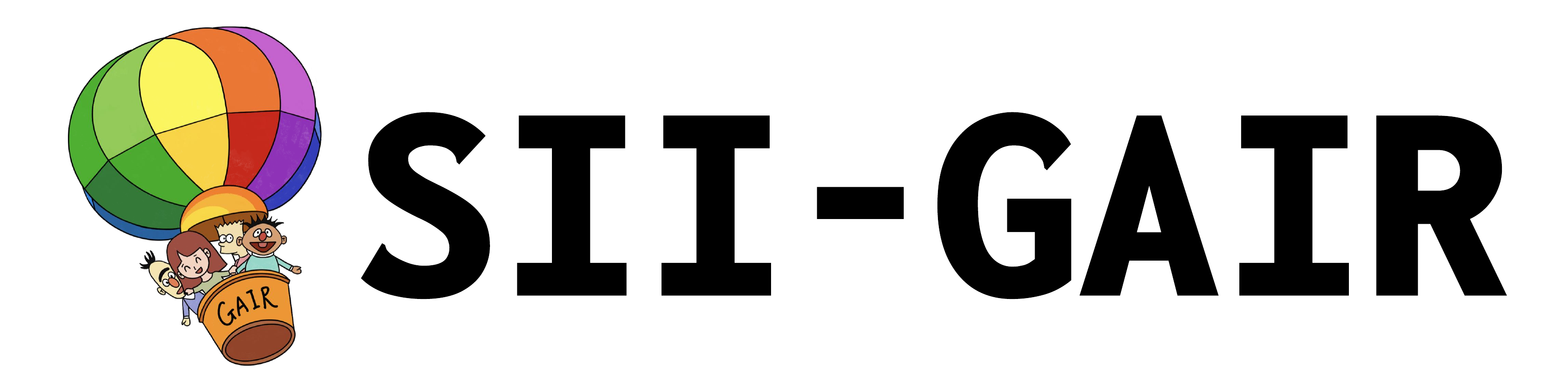}}%
}
\renewcommand{\headrulewidth}{0pt}
\setlength{\headsep}{2mm}

\renewcommand{\thefootnote}{}
\footnotetext{† Corresponding author.}
\vspace{-25pt}

{\centering
\href{https://github.com/sii-research/GAIR}{\raisebox{-.15em}{\includegraphics[height=1em]{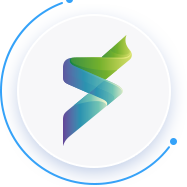}}\ SII Open Source:}
\quad \href{https://github.com/GAIR-NLP/daVinci-Dev}{\textcolor{black}\faGithub\ Code}
\quad \href{https://huggingface.co/GAIRdaVinci-Dev-72B}{\raisebox{-.15em}{\includegraphics[height=1em]{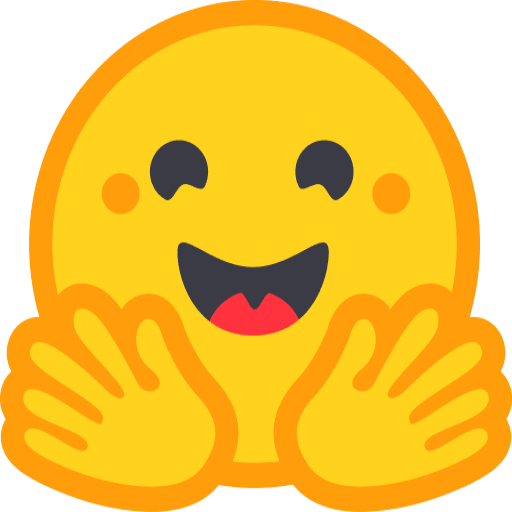}}\ Models}
\quad \href{https://huggingface.co/datasets/GAIR/daVinci-Dev}{{\textcolor{violet}\faDatabase}\ Datasets}
\par}

\vspace{20pt}


\newcommand{\tctx}{\tau^{\mathrm{ctx}}}
\newcommand{\tenv}{\tau^{\mathrm{env}}}

\newcommand{\Dataset}[2]{\mathcal{D}^{#1}\if\relax\detokenize{#2}\relax\else_{#2}\fi}

\newcommand{\Dctx}{\Dataset{\mathrm{ctx}}{}}
\newcommand{\Denv}{\Dataset{\mathrm{env}}{}}
\newcommand{\Dctxpy}{\Dataset{\mathrm{ctx}}{\text{py}}}
\newcommand{\Dctxgen}{\Dataset{\mathrm{ctx}}{\text{gen}}}
\newcommand{\Denvpass}{\Dataset{\mathrm{env}}{\text{pass}}}
\newcommand{\Denvfail}{\Dataset{\mathrm{env}}{\text{fail}}}
\newcommand{\Dswe}{\Dataset{\text{SWE-smith}}{}}

\vspace{-10px}
\begin{abstract}

Recently, the frontier of Large Language Model (LLM) capabilities has shifted from single-turn code generation to agentic software engineering—a paradigm where models autonomously navigate, edit, and test complex repositories.
While post-training methods have become the de facto approach for code agents, \emph{agentic mid-training}—mid-training (MT) on large-scale data that mirrors authentic agentic workflows—remains critically underexplored due to substantial resource requirements, despite offering a more scalable path to instilling foundational agentic behaviors than relying solely on expensive reinforcement learning.
A central challenge in realizing effective agentic mid-training is the distribution mismatch between static training data and the dynamic, feedback-rich environment of real development.
To address this, we present a systematic study of agentic mid-training, establishing both the data synthesis principles and training methodology for effective agent development at scale.
Central to our approach is \emph{agent-native data}—supervision comprising two complementary types of trajectories: \emph{contextually-native trajectories} that preserve the complete information flow an agent experiences, offering broad coverage and diversity; and \emph{environmentally-native trajectories} collected from executable repositories where observations stem from actual tool invocations and test executions, providing depth and interaction authenticity.
We verify the model’s agentic capabilities on \texttt{SWE-Bench Verified}.
We demonstrate our superiority over the previous open software engineering mid-training recipe \textsc{Kimi-Dev} under two post-training settings with an aligned base model and agentic scaffold, while using less than half mid-training tokens (73.1B).
Besides relative advantage, our best performing 32B and 72B models achieve \textbf{56.1\%} and \textbf{58.5\%} resolution rates, respectively, which are
state-of-the-art among open training recipes using agentic scaffolds under their model sizes, despite starting from non-coder \texttt{Qwen2.5-Base} base models.
Beyond these agentic capabilities, we also observe performance gains on general code generation and scientific benchmarks.
We plan to open-source a significant portion of our datasets, recipes, and model checkpoints—resources representing substantial computational investment typically unavailable to the broader community—to facilitate further research in this underexplored paradigm.

\end{abstract}

\newlength{\imgAw}\newlength{\imgAh}
\newlength{\imgBw}\newlength{\imgBh}

\settowidth{\imgAw}{\includegraphics{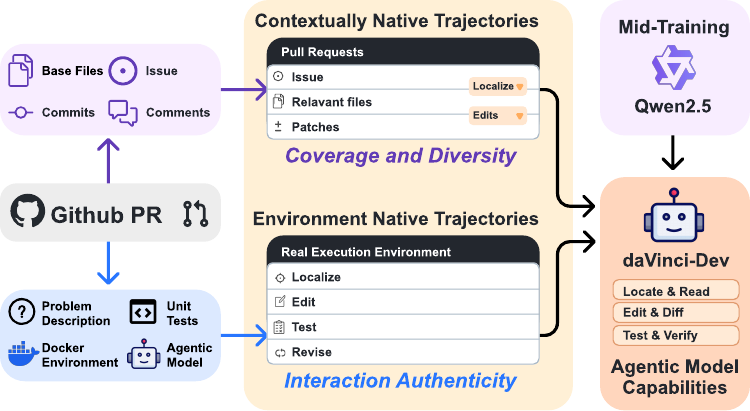}}
\settototalheight{\imgAh}{\includegraphics{figures/teaser_1.pdf}}
\settowidth{\imgBw}{\includegraphics{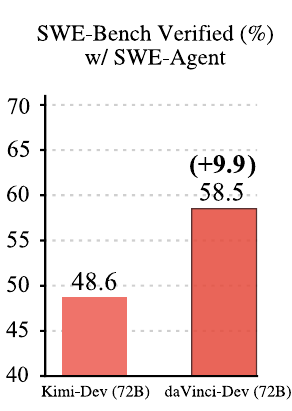}}
\settototalheight{\imgBh}{\includegraphics{figures/teaser_2.pdf}}

\pgfmathsetmacro{\aspectA}{\imgAw/\imgAh}
\pgfmathsetmacro{\aspectB}{\imgBw/\imgBh}
\pgfmathsetmacro{\fracA}{\aspectA/(\aspectA+\aspectB)*0.80}
\pgfmathsetmacro{\fracB}{\aspectB/(\aspectA+\aspectB)*0.80}

\vspace{-10px}
\begin{figure}[htbp]
    \centering
    \begin{subfigure}[b]{\fracA\linewidth}
        \includegraphics[width=\linewidth]{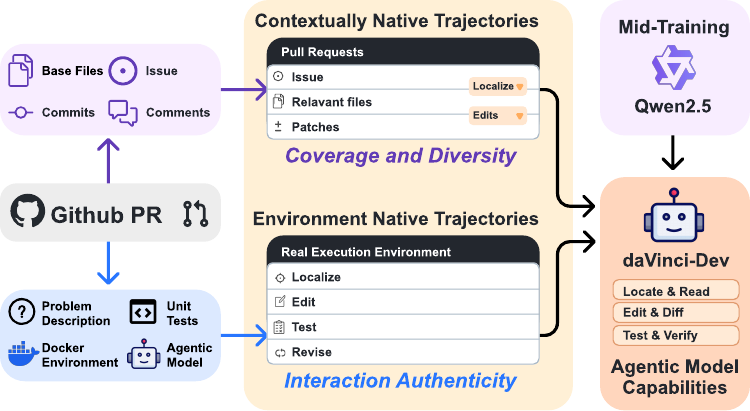}
        \label{fig:teaser_left}
    \end{subfigure}%
    \begin{subfigure}[b]{\fracB\linewidth}
        \includegraphics[width=\linewidth]{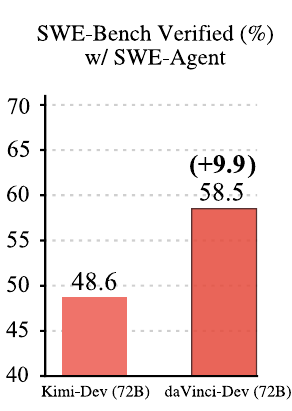}
        \label{fig:teaser_right}
    \end{subfigure}
    \vspace{-15px}
    \caption{Overview of our recipe. \textbf{Left:} We curate two complementary datasets using different elements of PRs. \textbf{Right:} Comparison of our best performance and the best performance of the previous open software engineering mid-training recipe \textsc{Kimi-Dev} on \textsc{SWE-Agent} scaffold. Detailed comparison is reported in Table \ref{tab:main_ablation}.}
    \label{fig:teaser}
\end{figure}

\vspace{-40px}
\newpage
\pagestyle{fancy}
\lhead{\rightmark}
\renewcommand{\headrulewidth}{0.7pt}
\setlength{\headsep}{5mm}
\clearpage

\newpage
\renewcommand{\thefootnote}{\arabic{footnote}}
\setcounter{footnote}{0}

\section{Introduction}
\label{sec:introduction}

The capabilities of code-generating large language models have rapidly expanded from synthesizing isolated functions~\citep{jain2024livecodebench, wang2024your} to tackling repository-level software engineering tasks~\citep{jimenez2023swe}. This shift toward \emph{agentic software engineering}~\citep{jimenez2023swe, badertdinov2025scaling,wu2025innovatorbench} reflects the demands of real-world development, where resolving issues requires code agents to autonomously and iteratively navigate complex codebases, understand cross-file dependencies, apply edits, and validate changes through test execution.
The dominant approach to building such code agents has centered on post-training: supervised fine-tuning (SFT) on curated trajectories~\citep{yang2025swesmithscalingdatasoftware, yang2025kimidevagentlesstrainingskill} followed by reinforcement learning (RL) from execution feedback~\citep{team2025kimi}. While effective, \textbf{the quantity and diversity are limited} for the repositories that can be used in this paradigm. Due to the unclear instructions in the repositories' README files and the resource limitation (e.g., GPUs), the number of repositories that can be transformed into executable environments is limited. Moreover, in most cases, only correct trajectories can be used in training, but high-quality agentic trajectories generated by expert human annotators are expensive to collect, while the sophisticated agent systems can only solve a small subset of issues, so most environments will also be filtered before the training process~\citep{yu2025dapoopensourcellmreinforcement}. Such a flaw constrains the learning dynamics and degrades performance during SFT or RL training. More fundamentally, post-training is constrained by the base model's intrinsic capacities, and certain agentic reasoning abilities may not be learnable through post-training alone~\citep{ye2025limo}.

This raises a natural question: \emph{Can we instill foundational agentic behaviors earlier in the training pipeline, during mid-training?} Mid-training (MT) on domain-specific data has proven transformative for specializing LLMs to domains like mathematics~\citep{lewkowycz2022solvingquantitativereasoningproblems, wang2025octothinkermidtrainingincentivizesreinforcement} and code~\citep{yang2025kimidevagentlesstrainingskill}. For agentic software engineering, mid-training offers a compelling value proposition: by exposing base models to massive-scale data that mirrors agentic iterations—file navigation, contextual edits, tool invocations, test-driven iterations—we can build stronger foundations that subsequent post-training can refine more efficiently. Yet despite this potential, \emph{agentic mid-training remains critically underexplored}.
Existing mid-training or pre-training efforts for code models~\citep{ yang2025kimidevagentlesstrainingskill} predominantly adopt a factorized approach: synthesizing isolated samples for atomic capabilities such as localization and editing, without the procedural context that an agent would encounter before exercising these capabilities.

In effect, we argue that existing MT data is not \emph{agent-native}: it does not preserve the action–observation loop structure that governs real development. We therefore identify the core challenge as a \textbf{distribution mismatch} between conventional training data and the dynamic reality of agentic deployment. Consider a typical GitHub Pull Request: while the commit history reveals \emph{what} files were changed, it obscures \emph{how} a developer (or agent) discovered those files, what context they examined before editing, and how test feedback shaped subsequent revisions. Training on such static snapshots—even at massive scale—leaves models unprepared for the sequential, interactive nature of real development workflows. An agent should not just learn to \emph{navigate} to the right location, \emph{retrieve} relevant context, \emph{generate} correct edits, \emph{apply} changes, and \emph{run} unit tests separately, but should coordinate these skills into a coherent, iterative problem-solving loop, where feedback from each step informs the next action.

To bridge this gap, we present the first systematic study of agentic mid-training for software engineering at scale. Our central thesis is that effective agentic mid-training requires \textbf{large-scale and diverse} \textbf{\emph{agent-native data}}—supervision that preserves the complete information flow and environmental dynamics an agent experiences during deployment. We formalize this through two complementary trajectory types:

\begin{itemize}[leftmargin=*, itemsep=3pt]
\item \textbf{Contextually-native trajectories}:
This type of trajectory emphasizes \textbf{coverage and diversity}. Any supervision instance that preserves the structure of a realistic engineering process can be included, regardless of whether it was produced through live execution, ensuring broad source repository coverage and diversity. Supervision is organized around full task-level action sequences, bundling localization steps (e.g., identifying relevant files) together with modification steps (e.g., applying edits). Edit actions can be conducted multiple times with interleaved textual reasoning to reflect a coherent software development process. This allows the dataset to capture a wide variety of valid contextual patterns and operational permutations.

\item \textbf{Environmentally-native trajectories}:
This type of trajectory prioritizes \textbf{interaction authenticity} while also considering quantity. Only trajectories generated through actual interactions with a real development environment are eligible for inclusion. These trajectories record genuine observations—tool invocations, test executions, runtime errors, and scaffold system feedback—rather than simulated or retrospectively constructed observations. We do not apply any filter strategy, so that the quantity of such trajectories can be much larger than the ones in the SFT stage. This exposes models to the dynamic feedback loops inherent in real development.
\end{itemize}

We materialize these principles through a large-scale data synthesis effort that leverages \emph{different elements} from GitHub Pull Requests to construct two complementary data types, as illustrated in Figure~\ref{fig:teaser}. First, we curate a \textbf{68.6B-token contextually-native trajectories} ($\Dctx$) using base files and commits, carefully reconstructing the procedural process behind each code change: which context the developer likely examined (related issue and base file content), and how they iteratively refined their solution (temporal commits). This transforms static diffs into contextually-rich trajectories that preserve the natural coupling between navigation and editing, providing \emph{broad coverage and diversity} across repositories and languages. Second, we construct \textbf{3.1B-token environmentally-native trajectories} ($\Denv$) from PR-derived software engineering tasks using their Docker environments and unit tests, generating agentic rollouts where our agent interacts with real build systems, test suites, and linters, collecting observations from actual tool outputs. This provides \emph{depth and authenticity} through genuine execution feedback that cannot be retrospectively reconstructed.

Evaluating our models on \texttt{SWE-Bench Verified}, we surpass the previous state-of-the-art open MT recipe, \textsc{Kimi-Dev}, under two post-training settings with an aligned base model and agentic scaffold while reducing the mid-training corpus size by over 50\% (73.1B vs $\sim$150B tokens). Our best performing 32B and 72B models reach resolution rates of \textbf{56.1\%} and \textbf{58.5\%}, respectively. These scores represent the highest performance among open training recipes using agentic scaffolds for their respective model sizes, a significant feat given our initialization from \texttt{Qwen2.5-Base} models instead of newer or code-focused base models. Beyond agentic workflows, this regimen also confers broad benefits, improving performance on scientific and general code generation tasks as detailed in Table~\ref{tab:generalization_benchmarks}.

To conclude, we make the following contributions:

\begin{itemize}[leftmargin=*, itemsep=3pt]
\item We formulate \textbf{agentic mid-training} and introduce \textbf{agent-native data} as supervision that preserves the information flow of real software engineering. We build large-scale agent-native corpora from public software development traces, including a \textbf{68.6B-token} contextually-native corpus and a \textbf{3.1B-token} set of environmentally-native rollouts, and provide a practical training recipe that leverages them.
\item We demonstrate consistent gains on agentic software engineering brought by our agentic mid-training recipe across post-training schemes and model sizes, and provide analysis of robustness, scalability, and generalization. We plan to release the data construction code, training configurations, and a substantial portion of the resulting artifacts (e.g., curated datasets and checkpoints) where permitted.
\end{itemize}

\section{Background and Problem Setup}
\label{sec:background}

\subsection{Agentic Software Engineering Tasks}
\label{sec:prelim_setup}

We formalize an agentic software engineering task as a tuple $(\mathcal{R}, q, \mathcal{E})$, where $\mathcal{R}$ is a repository state, $q$ is a natural language problem description (e.g., bug report, issue), and $\mathcal{E}$ is an evaluation oracle (typically a test suite). Unlike single-turn generation where all necessary context is provided upfront, agentic tasks require multi-step interaction. At each step $t$, the agent selects an action based on the conversation history and receives an observation from an observation generator:
\begin{align*}
 a_t &\sim \pi_\theta(a \mid h_{t-1}, q) \quad \text{(action selection)} \\
 o_t &\sim \text{Obs}(a_t, \mathcal{R}) \quad \text{(observation)}
\end{align*}
where $h_{t-1} = \{(a_1, o_1), \ldots, (a_{t-1}, o_{t-1})\}$ accumulates prior interactions.

Actions correspond to tool calls such as searching for files, reading code, applying edits, or running tests, while observations return concrete outputs like file contents, compiler errors, or test results. This interaction is necessary because the agent initially does not know which parts of the codebase (potentially thousands of files) are relevant to the issue, and must iteratively refine its solution based on feedback from the evaluation oracle.

While the exact sequence varies by task complexity, a typical development workflow follows the pattern: \texttt{localize} (identifying relevant files) $\rightarrow$ \texttt{read} (understanding code context) $\rightarrow$ \texttt{edit} (applying modifications) $\rightarrow$ \texttt{test} (validating changes) $\rightarrow$ \texttt{revise} (refining based on feedback). This structure reflects common agent implementations~\citep{yang2025swesmithscalingdatasoftware} and mirrors natural software development practices, though agents may repeat or interleave these steps as needed.

\begin{figure}[t]
    \centering
    \includegraphics[width=\linewidth]{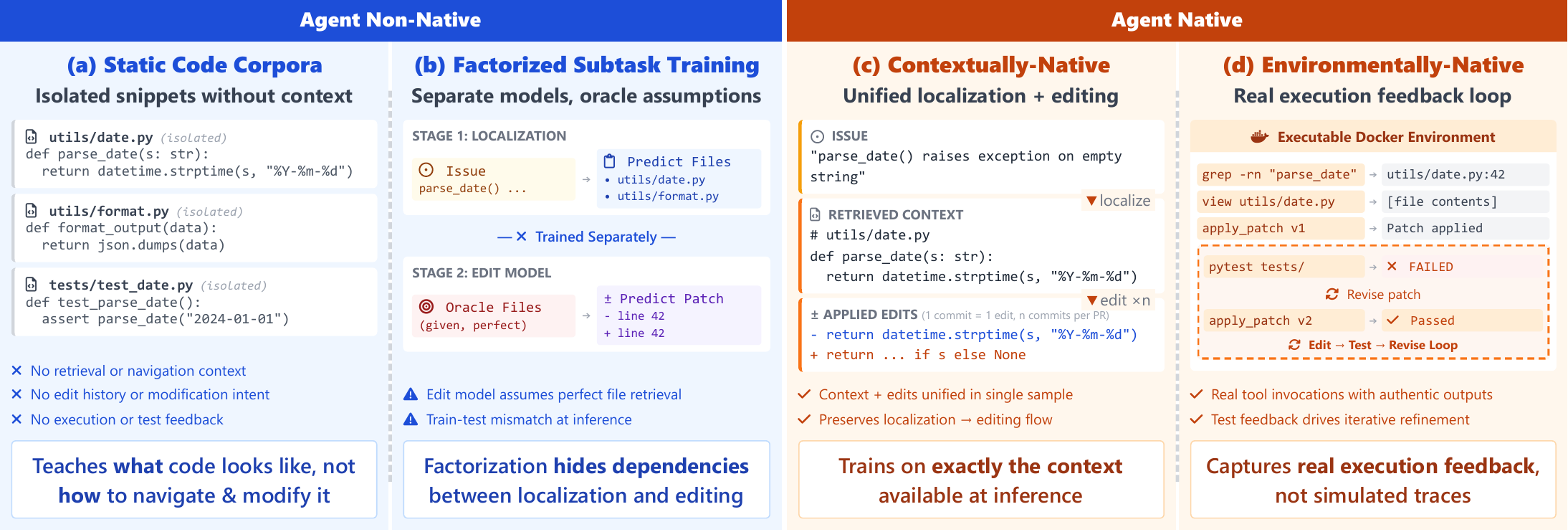}
    \caption{Comparison of training data paradigms. (a) Traditional code pre-training uses isolated static files. (b) Factorized approaches train subtasks separately, creating train-test mismatch. (c) Our contextually-native PRs bundle retrieval context with sequential edit trajectory. (d) Our environmentally-native trajectories capture real execution feedback loops.}
    \label{fig:our_data}
\end{figure}

This complete sequence is an \textbf{agent trajectory} $\tau = (q, \mathcal{R}, \{(a_i, o_i)\}_{i=1}^T, y)$, where $y \in \{0,1\}$ indicates whether the trajectory is successful under its supervision source.

\subsection{The Distribution Mismatch Problem}
\label{sec:distribution_mismatch}

As formalized in Section~\ref{sec:prelim_setup}, agentic software engineering requires multi-step interaction. However, traditional training data predominantly consists of static, completed artifacts that bear no resemblance to this interactive process (Figure~\ref{fig:our_data}a). Models are exposed to final outcomes—complete code files, merged commits, finished implementations—without the sequential action-observation pairs $(a_t, o_t)$ that agents experience at deployment. This creates a critical distribution mismatch: training data shows \emph{what} was ultimately produced, but deployment requires agents to learn \emph{how} to construct solutions through the dynamic workflow of localization, reading, editing, testing, and revision.

Moreover, even when workflow structure is preserved, training data typically shows only successful final states, omitting the validation failures, error messages, and iterative refinements that emerge from actual environment interaction. Models trained on such incomplete supervision must learn to orchestrate complete workflows and respond to execution feedback during post-training, rather than internalizing these coordination patterns as foundational behaviors.

\section{Agent-Native Data: Design and Synthesis}
\label{sec:dataset_construction}

To address the distribution mismatch between static training data and interactive deployment, we construct agent-native data—supervision that preserves the complete action-observation trajectories and environmental feedback agents experience during real problem-solving. Specifically, we construct two complementary types of agent-native trajectories: \textbf{contextually-native trajectories}, which reconstruct complete workflows from GitHub Pull Requests to preserve the full development context, and \textbf{environmentally-native trajectories}, which capture authentic execution feedback through agent rollouts in real executable environments. The combination ensures both breadth (diverse workflow patterns at scale) and depth (authentic execution dynamics), addressing the distribution mismatch from complementary angles.

\subsection{Contextually-Native Trajectories}
\label{sec:contextually_native}

\subsubsection{Design Rationale}
To construct contextually-native trajectories, we leverage GitHub Pull Requests (PRs) as the base data source. PRs naturally connect problem specifications (issues) to solutions (code changes) with validation signals (tests, reviews), making them well-suited for reconstructing development workflows. The key design principle is bundling complete context: rather than factorizing PRs into independent localization and editing tasks~\citep{yang2025kimidevagentlesstrainingskill} (Figure ~\ref{fig:our_data}b), we keep all relevant information together—issue description, relevant repository files, and modifications—in a single training sample. This preserves the causal flow agents experience at deployment, where editing decisions must be conditioned on the context gathered during localization.

\subsubsection{Construction Pipeline}

\paragraph{Data Sources.}

We construct contextually-native trajectories from two complementary subsets: $\Dctxgen$ (``general'') provides broad coverage of software engineering patterns across diverse languages and frameworks by drawing from highly-starred repositories, while $\Dctxpy$ (``Python'') ensures strong alignment with software engineering benchmarks (e.g., \texttt{SWE-Bench Verified}) through focused coverage of Python development. The two subsets partially overlap in Python repositories but serve complementary purposes: $\Dctxgen$ establishes cross-language understanding, while $\Dctxpy$ ensures alignment with target evaluation tasks.

\paragraph{Collection.}

We collect pull requests through GitHub REST\footnote{REST API: \url{https://docs.github.com/en/rest}} and GraphQL APIs\footnote{GraphQL API: \url{https://docs.github.com/en/graphql}}. For each repository, we obtain pull request metadata and selectively query additional endpoints for detailed content, including linked issue descriptions (if exist), relevant file contents at the base commit, and the full commit sequence with corresponding diffs. We determine relevant files deterministically by querying the net diff between base and head commits. To ensure correctness, we align file contents and patches with the parent of the first PR commit, rather than using the base commit recorded in PR metadata, which may not reflect the actual codebase state when the PR was created.

\paragraph{Filtering.}

We apply multi-level filtering criteria to ensure data quality while maintaining coverage. (1) At the repository level, $\Dctxgen$ selects from the top 10,000 most-starred repositories across all languages\footnote{In the future, we will scale our approach to more repositories.}. $\Dctxpy$ focuses on public Python repositories (\texttt{language=Python} in metadata API) with at least 5 stars and not archived. The relaxed star threshold for $\Dctxpy$ balances repository diversity with quality standards. (2) At the pull request level, both subsets retain only merged PRs and exclude bot-created PRs. For $\Dctxpy$, we additionally require modifications to be done only in Python source or documentation files, with the number of changed Python files between 1 and 5. Six million out of thirteen million pull requests pass the $\Dctxpy$ pull request level filters. These constraints ensure focused, manageable tasks suitable for agent learning.

\begin{figure}[!t]
    \centering
    \includegraphics[width=1\linewidth]{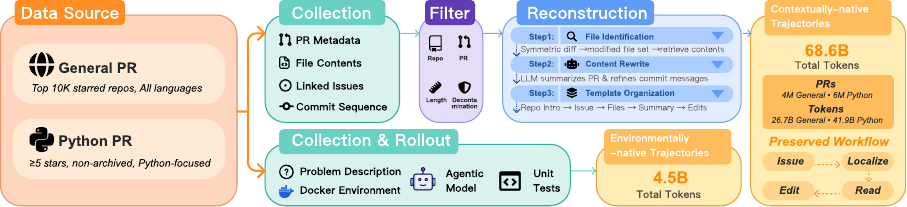}
    \caption{Overview of our dataset generation pipeline.}
    \label{fig:pipeline}
\end{figure}
\paragraph{Reconstruction.}

For each filtered PR, we reconstruct an agent-native training sample through the following process: (1) \emph{Content enhancement.} We use \texttt{Qwen3-235B-A22B-Instruct-2507}~\citep{yang2025qwen3} to generate two types of enhancements. First, we create a concise summary of the overall PR that captures its intent and main changes. Second, since some commit messages are terse or uninformative (e.g., ``fix'', ``update''), we refine them into more descriptive summaries that explain what each commit accomplishes. Detailed prompts are provided in Appendix~\ref{app:llm_prompts_pr_rendering}. (2) \emph{Relevant file identification.} We identify relevant files through reverse engineering: analyzing symmetric diffs across all commits to extract the set of modified files, then retrieving their complete contents at the base commit. (3) \emph{Template organization.} We organize all extracted information into clearly delineated sections: Repository Context, Issue (when available), Pull Request, Relevant Files Found (complete file contents), LLM-generated Summary, and Edits. The Edits section contains the code modifications: for PRs with multiple commits, we concatenate them in temporal order, with each refined commit message followed by its associated code changes. The two subsets use different structural formats: $\Dctxgen$ adopts XML-like tags with traditional patch diffs and additionally includes developer comments and reviews, while $\Dctxpy$ uses Markdown structure with search-and-replace blocks that more directly represent agent editing actions. See Appendix~\ref{app:pr_data_template} for detailed format specifications and examples.

This organization mirrors the workflow: relevant file paths simulates the ``localize'' phase, file contents represent the ``read'' phase, edits represent the ``edit'' phase, and LLM-generated contents serve as textual reasoning in between.

We apply standard post-processing to ensure training efficiency and evaluation integrity. For length filtering, we discard samples exceeding 32k tokens, which retains over 90\% of Python pull requests while improving training efficiency. For decontamination, we remove all pull requests from repositories included in \texttt{SWE-Bench Verified} to prevent data leakage and ensure fair evaluation.

\subsubsection{Corpus Statistics}

After applying the pipeline described, we obtain two complementary subsets. The general subset $\Dctxgen$ (26.7B tokens) provides broad software engineering coverage across diverse languages and tooling patterns, drawn from 4 million PRs in the top 10,000 most-starred repositories. The Python subset $\Dctxpy$ (41.9B tokens), comprising 6 million PRs from $7.4 \times 10^5$ repositories, focuses on Python development to ensure alignment with benchmarks. Together, $\Dctx = \Dctxgen \cup \Dctxpy$ contains 68.6B tokens of contextually-native trajectories and preserves complete development workflows across diverse repositories and scenarios.

\subsection{Environmentally-Native Trajectories}
\label{sec:environmentally_native}

\noindent\textbf{Notation.} We denote the environmentally-native dataset as $\Denv$, consisting of trajectories $\tenv$. We further split it into $\Denvpass$ and $\Denvfail$ based on the final test outcome.

\subsubsection{Design Rationale}
While contextually-native trajectories provide agent-like \emph{structure}, they lack agent-like \emph{dynamics}: the model never observes the iterative feedback loop (edit $\rightarrow$ test $\rightarrow$ revise) that characterizes real agentic coding in practice. To close this gap, we curate environmentally-native trajectories---collected by running a capable agent in real executable development environments with authentic test feedback.

Our approach contrasts with trajectories in simulated or synthetic environments~\citep{yang2025kimidevagentlesstrainingskill} where codebase navigation is read-only or test execution is unavailable during rollout. While such approaches produce trajectories with agentic format, they lack agentic feedback—the model never observes how its edits affect test outcomes or how error messages guide revisions. Environmentally-native trajectories preserve this critical feedback loop by recording actual agent-environment interactions: tool invocations, test executions, and the resulting observations (test outputs, error messages, runtime feedback).

\subsubsection{Construction Pipeline}

In order to ensure authenticity, we choose to derive executable environments from real GitHub pull requests, rather than artificially constructed ones. Therefore, we construct our agentic rollout environments following the methodology established in \textsc{SWE-rebench}~\citep{badertdinov2025scaling}. We build a Docker image for each task that reproduces the repository state at a specific commit, alongside unit tests from the actual codebase, and infrastructure to execute tool calls (file edits, shell commands, test runs). Then we deploy \texttt{GLM-4.6}~\citep{5team2025glm45agenticreasoningcoding} within the \textsc{SWE-Agent} framework~\citep{yang2025swesmithscalingdatasoftware}. For each task, we generate up to 4 rollouts, recording the complete action-observation sequences: the agent's actions and the environment's responses (file contents, search results, test outcomes, error messages).

After discarding trajectories exceeding 128k tokens, we classify the remaining trajectories based on final test outcomes into passing trajectories (all tests pass) and non-passing trajectories (tests fail). Both types provide valuable learning signals: passing trajectories demonstrate complete problem-solving cycles, while non-passing trajectories capture realistic debugging scenarios with error feedback.

\subsubsection{Corpus Statistics}

After filtering, we obtain two types of environmentally-native trajectories: $1.85 \times 10^{4}$ \textbf{passing trajectories} $\Denvpass$ (0.7B tokens) where all tests pass, and $5.55 \times 10^{4}$ \textbf{non-passing trajectories} $\Denvfail$ (2.4B tokens) with test failures, totaling approximately $7.4 \times 10^{4}$ trajectories and 3.1B tokens. The corpus features the authentic execution feedback—test results, runtime errors and iterative refinements that complement the workflow structure learned from PR data.

\section{Experiments}
\label{sec:experiments}

\subsection{Training Pipeline Terminology}
\label{sec:training_pipeline}

We clarify our position within the standard LLM development pipeline:

\paragraph{Pre-training.} Large-scale next-token prediction on diverse corpora.

\paragraph{Mid-training (MT).} An intermediate stage that shifts capability distribution by training on curated domain data at scale~\citep{wang2025octothinkermidtrainingincentivizesreinforcement}. Unlike fine-tuning (which teaches specific behaviors), mid-training operates at the knowledge level.

\paragraph{Post-training.} Supervised fine-tuning (SFT) on demonstrations and/or reinforcement learning.

\subsection{Experimental Setup}
\label{sec:exp_setup}

\paragraph{Base model.}
Unless otherwise specified, we start from base model  \texttt{Qwen2.5-72B-Base} and \texttt{Qwen2.5-32B-Base}.

\paragraph{Evaluation.}
We evaluate on \texttt{SWE-Bench Verified} using \textsc{SWE-Agent} (temperature 0, 128k context and 100 steps) and report Pass@1, averaged across 4 runs. We manually fix a small number of test cases where the provided ground truth patch cannot pass due to various reasons.

\paragraph{Training stages.}
We consider two stages:
(i) \textbf{mid-training (MT)} on large-scale unlabeled corpora (PR data and/or trajectories), and
(ii) \textbf{supervised fine-tuning (SFT)} on agentic trajectories.
The training configuration is detailed in \S\ref{app:training_details}.

\paragraph{Data components.}
We use three main data sources. For compactness in tables, we denote datasets with symbols (defined in Section~\ref{sec:prelim_setup}):
\begin{itemize}
    \item For \textbf{contextually-native trajectories} ($\Dctx$), we transform GitHub pull requests into the structured format described in \S\ref{sec:contextually_native}. In this setting, we use two subsets:
    \begin{itemize}
        \item $\Dctxpy$ (41.9B): a Python-focused subset for alignment with software engineering benchmarks.
        \item $\Dctxgen$ (26.7B): a general subset drawn from most-starred repositories across all languages.
        \item $\Dctx$ (68.6B): $\Dctxgen \cup \Dctxpy$.
    \end{itemize}
    \item For \textbf{environmentally-native trajectories} ($\Denv$), we collect rollouts by running \textsc{SWE-Agent} with \texttt{GLM-4.6} in executable Docker environments derived from real GitHub pull requests, forming $\Denv$ (3.1B raw tokens; $\sim$4.5B effective tokens). We upsample $\Denvpass$ by $3\times$ during training.
    \item For data to activate the model after mid-training, we may use:
    \begin{itemize}
        \item $\Denvpass$ (0.7B): subset of $\Denv$ that pass the unit tests
        \item $\Dswe$ (0.11B tokens): a public set of \textsc{SWE-Agent} trajectories released by \citet{yang2025swesmithscalingdatasoftware} (mostly generated with Claude 3.7 Sonnet), which we use as an external SFT baseline.
    \end{itemize}
\end{itemize}

\paragraph{Baselines.}
For Kimi-Dev comparisons, we quote results from \citet{yang2025kimidevagentlesstrainingskill} where applicable, and match our SFT dataset $\Dswe$ and parameters (\S\ref{app:training_details}) close to theirs. For experiments requiring downstream SFT on $\Denvpass$ (Table~\ref{tab:main_ablation}), we utilize the official \texttt{Kimi-Dev-72B} checkpoint as the starting point, as their pre-RL mid-training checkpoint is not publicly available.

\subsection{Mid-Training Provides Robust Gains}
\label{sec:exp_mt_value}

\begin{table}[t]
\centering
\small
\begin{tabular}{l c c c c}
\toprule
\textbf{Model / Variant} & \makecell{\textbf{Mid-training}\\\textbf{Data}} & \makecell{\textbf{Post-training}\\\textbf{Data}} & \makecell{\textbf{Post-training}\\\textbf{Method}} & \textbf{SWE-V} \\
\midrule
\multicolumn{5}{l}{\textit{Qwen 2.5 32B Series}} \\
\rowcolor{tableblue!25}
Baseline (Weak SFT)\textsuperscript{\textdagger} & - & $\Dswe$ & SFT & 34.8 \\
\rowcolor{tableblue!25}
Baseline (Strong SFT)\textsuperscript{\textdagger} & - & $\Denvpass$ & SFT & 53.0 \\
\rowcolor{tableorange!60}
Ours (Weak SFT) & $\Dctx$ & $\Dswe$ & SFT & 39.5 \\
\rowcolor{tableorange!60}
Ours (Strong SFT) & $\Dctx$ & $\Denvpass$ & SFT & 54.1 \\
\rowcolor{tableorange!60}
\textbf{Ours (daVinci-Dev-32B)} & \textbf{$\Dctx+\Denv$} & \textbf{$\Denvpass$} & \textbf{SFT} & \textbf{56.1} \\
\addlinespace[2pt]
\midrule
\multicolumn{5}{l}{\textit{Qwen 2.5 72B Series}} \\
\rowcolor{tableblue!30}
Baseline (Weak SFT)\textsuperscript{\textdagger} & - & $\Dswe$ & SFT & 38.0 \\
\rowcolor{tableblue!30}
Baseline (Strong SFT)\textsuperscript{\textdagger} & - & $\Denvpass$ & SFT & 56.6 \\
\rowcolor{tableblue!30}
Kimi-Dev~\citep{yang2025kimidevagentlesstrainingskill} & $\Dataset{\text{AgentlessMT}}{}$ & $\Dataset{\text{SWE-smith}}{}$ & SFT & $\approx$46.0\textsuperscript{\textdaggerdbl} \\
\rowcolor{tableblue!30}
Kimi-Dev~\citep{yang2025kimidevagentlesstrainingskill} & $\Dataset{\text{AgentlessMT}}{}$ & $\Dataset{\text{AgentlessRL}}{}$ + $\Dataset{\text{SWE-smith}}{}$ & SFT+RL & 48.6 \\
\rowcolor{tableblue!30}
Kimi-Dev~\citep{yang2025kimidevagentlesstrainingskill}\textsuperscript{\textdagger} & $\Dataset{\text{AgentlessMT}}{}$ & $\Dataset{\text{AgentlessRL}}{}$ + $\Dataset{\mathrm{env}}{\text{pass}}$ & SFT+RL & 56.2 \\
\rowcolor{tableorange!60}
Ours (Weak SFT) & $\Dctx$ & $\Dswe$ & SFT & 46.4 \\
\rowcolor{tableorange!60}
Ours (Strong SFT) & $\Dctx$ & $\Denvpass$ & SFT & 58.2 \\
\rowcolor{tableorange!60}
\textbf{Ours (daVinci-Dev-72B)} & \textbf{$\Dctx+\Denv$} & \textbf{$\Denvpass$} & \textbf{SFT} & \textbf{58.5} \\
\bottomrule
\end{tabular}
\caption{Ablations and mid-training comparisons on \texttt{SWE-Bench Verified} (SWE-V). Our agentic mid-training on contextually-native trajectories ($\Dctx$) and environmentally-native trajectories ($\Denv$) consistently improves downstream performance, and is competitive with or surpasses prior mid-training recipes. All results use \textsc{SWE-Agent} for evaluation.
\textsuperscript{\textdagger}Trained and tested using our infrastructure. \textsuperscript{\textdaggerdbl}Estimated from Figure 5 in \citet{yang2025kimidevagentlesstrainingskill}.}
\label{tab:main_ablation}
\end{table}

Our most important finding is that agent-native mid-training improves performance even when strong trajectory SFT already yields competitive results. For robustness we also validate this across two SFT regimes and against the strongest prior MT recipe (with best effort). The comparison results are shown in Table~\ref{tab:main_ablation}.

\paragraph{Robustness across SFT regimes.}
On the 72B model, our MT consistently boosts performance.
With weak SFT, we improve from 38.0\% to 46.4\% with only $\Dctx$ MT, matching the Kimi-Dev baseline despite using fewer than half the tokens (68.6B vs.\ 150B) and no synthetic reasoning data.
With strong SFT, we reach \textbf{58.2\%} with only $\Dctx$ MT, outperforming the RL-tuned Kimi-Dev checkpoint and SFT-only baseline. This indicates that our contextually-native representation—bundling file context and edits—successfully bridges the gap between pre-training and agentic fine-tuning.
With $\Dctx+\Denv$ MT, our performance further increases to our strongest result \textbf{58.5\%}, showing that adding environmentally-native trajectories to MT enables the model to internalize the dynamics of the execution environment.

\paragraph{Robustness across scales.}
The benefits of our MT recipes transfer effectively from 72B to the smaller 32B model. On the 32B scale, $\Dctx$ MT improves the weak SFT baseline by 4.7\% and the strong SFT baseline by 1.1\%, and $\Dctx+\Denv$ MT continues to deliver the best performance \textbf{56.1\%}, 3.1\% above the strong SFT baseline. This confirms that the effectiveness of contextually-native trajectories and environmentally native trajectories is not specific to a single model capacity.

\subsection{Comparison with Open Recipes}
\label{sec:exp_comparison}

\begin{table}[t]
\centering
\small
\begin{tabular}{l c c c c c}
\toprule
\textbf{Model} & \makecell{\textbf{Base}\\\textbf{or Inst.}} & \makecell{\textbf{Mid-}\\\textbf{training}} & \makecell{\textbf{Post-training}\\\textbf{Method}} & \textbf{Scaffold} & \textbf{SWE-V} \\
\midrule
\multicolumn{6}{l}{\textit{Qwen 2.5 32B Coder Series}} \\
R2EGym-Agent~\citep{jain2025r2egymproceduralenvironmentshybrid} & Base & No & SFT & R2E-Gym & 34.4 \\
Openhands-LM~\citep{wang2025openhands} & Inst. & No & SFT  & OpenHands & 37.2 \\
SWE-Agent-LM~\citep{yang2025swesmithscalingdatasoftware} & Inst. & No & SFT & SWE-Agent & 40.2 \\
SWE-Mirror-LM~\citep{wang2025swemirrorscalingissueresolvingdatasets} & Inst. & No & SFT & MOpenHands & 52.2 \\
Skywork-SWE~\citep{zeng2025skyworksweunveilingdatascaling} & Inst. & No & SFT & OpenHands & 38.0 \\
SWE-Dev~\citep{wang2025swedevbuildingsoftwareengineering}  & Inst. & No & SFT+RL & OpenHands & 36.6 \\
\midrule
\multicolumn{6}{l}{\textit{Qwen 3 32B Series}} \\
DeepSWE-Preview~\citep{luo2025deepswe}  & Inst. & No & RL & OpenHands & 42.2 \\
FrogBoss~\citep{sonwane2025bugpilotcomplexbuggeneration}  & Inst. & No & SFT & SWE Agent & 54.6 \\
SWE-Lego-Qwen3-32B~\citep{tao2026swelegopushinglimitssupervised}  & Inst. & No & SFT & OpenHands & 52.6 \\
\midrule
\multicolumn{6}{l}{\textit{Qwen 2.5 32B Series}} \\
\textbf{daVinci-Dev-32B (Ours)} & \textbf{Base} & \textbf{Yes} & \textbf{SFT} & \textbf{SWE-Agent} & \textbf{56.1} \\
\midrule
\multicolumn{6}{l}{\textit{Qwen 2.5 72B Series}} \\
Kimi-Dev~\citep{yang2025kimidevagentlesstrainingskill} & Base & Yes & SFT+RL & SWE-Agent & 48.6 \\
\textbf{daVinci-Dev-72B (Ours)} & \textbf{Base} & \textbf{Yes} & \textbf{SFT} & \textbf{SWE-Agent} & \textbf{58.5} \\
\bottomrule
\end{tabular}
\caption{Comparison with representative methods on \texttt{SWE-Bench Verified} (SWE-V). We include representative works with agentic scaffolds.}
\label{tab:main_comparison}
\end{table}

We compare our full recipe against representative open methods on \texttt{SWE-Bench Verified} based on the \texttt{Qwen2.5} model family and use agentic scaffolds. Table~\ref{tab:main_comparison} presents the results.

\paragraph{Results.}
Within the 72B scale, our \texttt{daVinci-Dev-72B} achieves \textbf{58.5\%}, surpassing 48.6\% for Kimi-Dev using the same base model and agentic scaffold. At 32B scale, \texttt{daVinci-Dev-32B} achieves \textbf{56.1\%}, which is state-of-the-art among open training recipes at this scale using agentic scaffolds, despite the fact that prior work uses \texttt{Qwen2.5-Coder-32B} series or \texttt{Qwen3-32B} while our method starts from non-coder \texttt{Qwen2.5-32B-Base}.

\subsection{Generalization Beyond SWE Tasks}
\label{sec:exp_generalization}

\definecolor{deltacolor}{HTML}{FF8C00}
\newcommand{\deltac}[1]{\textcolor{deltacolor}{#1}}

\begin{table*}[ht]
\centering
\small
\begin{tabular}{l ccc ccc}
\toprule
& \multicolumn{3}{c}{\textbf{Qwen2.5-32B}} & \multicolumn{3}{c}{\textbf{Qwen2.5-72B}} \\
\cmidrule(lr){2-4} \cmidrule(lr){5-7}
\textbf{Benchmark} & \textbf{Base} & \textbf{MT Mix} & \textbf{$\Delta$} & \textbf{Base} & \textbf{MT Mix} & \textbf{$\Delta$} \\
\midrule
\multicolumn{7}{l}{\textit{Scientific Benchmarks}} \\
GPQA-Main & 38.17 & 38.84 & \deltac{+0.67} & 43.30 & 44.87 & \deltac{+1.57} \\
SuperGPQA & 33.85 & 35.94 & \deltac{+2.09} & 37.76 & 39.27 & \deltac{+1.51} \\
SciBench & 18.46 & 20.49 & \deltac{+2.03} & 19.33 & 19.77 & \deltac{+0.44} \\
\midrule
\multicolumn{7}{l}{\textit{Code Benchmarks}} \\
HumanEval & 58.16 & 81.42 & \deltac{+23.26} & 64.27 & 76.73 & \deltac{+12.46} \\
EvalPlus & 50.13 & 71.31 & \deltac{+21.18} & 56.04 & 69.45 & \deltac{+13.41} \\
DS-1000 & 12.2 & 21.2 & \deltac{+9.0} & 21.4 & 24.7 & \deltac{+3.3} \\
\bottomrule
\end{tabular}
\caption{Generalization performance on scientific and code benchmarks. We report the base model performance and the impact of our MT stages. \textbf{MT Mix} refers to the model trained on $\Dctxpy$ + $\Denv$.
}
\label{tab:generalization_benchmarks}
\end{table*}

While our agentic mid-training is specialized for software engineering, we investigate whether the agentic capabilities acquired from processing Pull Requests and execution trajectories transfer to broader domains requiring complex logic. We focus our evaluation on two distinct categories: standard code generation and rigorous scientific reasoning. In this experiment we choose a clean single stage MT recipe with $\Dctxpy$ + $\Denv$ as dataset.

As reported in Table~\ref{tab:generalization_benchmarks}, our model demonstrates strong generalization performance, consistently surpassing the base models across both 32B and 72B scales. In code generation, we observe substantial gains on HumanEval~\citep{chen2021codex} and EvalPlus~\citep{evalplus} (after decontamination following \texttt{XCoder}~\citep{wang2024codellmsperformempowering}), confirming that our data improves fundamental coding proficiency. More notably, we observe transfer learning to scientific benchmarks such as GPQA~\citep{rein2023gpqa} and SciBench~\citep{wang2024scibench}. These tasks, which demand expert-level domain knowledge and multi-step reasoning, benefit from the decision-making patterns inherent in our agentic mid-training. This suggests that the logic required for autonomous software engineering fosters fundamental reasoning skills that generalize beyond code.

\section{Analysis}
\label{sec:analysis}

In this section, we analyze the factors contributing to the effectiveness of agentic mid-training. We first examine the efficiency and information density of contextually-native data, then explore the synergistic relationship between our two data types, and finally discuss the scalability of this paradigm.

\subsection{High Information Density and Efficiency}
\label{sec:analysis_efficiency}

A key advantage of our approach is token efficiency. Kimi-Dev's recipe involves 70B tokens directly derived from PR plus 20B synthetic trajectory/CoT tokens upsampled 4 times, totaling $\sim$\textbf{150B tokens}. In contrast, our \textbf{68.6B tokens} $\Dctx$ MT stage consistently outperforms Kimi-Dev as shown in section~\ref{sec:exp_mt_value}, and performance further grows with additional \textbf{4.5B effective tokens} $\Denv$ added to MT training stage. This efficiency stems from our contextually-native representation being closer to software engineering agent's test  distribution compared to factorized approaches, and our environmentally-native trajectories being more authentic than simulated trajectories.

\subsection{Synergy: contextually-native data amplifies trajectory learning}
\label{sec:analysis_synergy}

While environmentally-native trajectories provide the correct format for agentic interaction, we find they are insufficient for generalization when used in isolation. Table~\ref{tab:synergy} presents an ablation study on the composition of MT data across both 32B and 72B model scales.

\begin{table}[h]
\centering
\small
\setlength{\tabcolsep}{4pt}
\begin{tabular}{l c c c c}
\toprule
 & & & \multicolumn{2}{c}{\textbf{SWE-Verified}} \\
\textbf{MT Data Composition} & \textbf{Tokens} & \textbf{SFT Data} & \textbf{32B Base} & \textbf{72B Base} \\
\midrule
\multicolumn{5}{l}{\textit{Ablation: Trajectories vs. PR (Zero-shot / No SFT)}} \\
$\Denv$ & 4.5B & -- & 43.7 & 47.1 \\
$\Denv$ + $\Dctxpy$ & 46.4B & -- & \textbf{49.9} & \textbf{54.8} \\
\midrule
\multicolumn{5}{l}{\textit{Impact of MT Composition on SFT}} \\
$\Dctxpy$ & 41.9B & $\Denvpass$ & 52.9 & 56.5 \\
$\Denv$ + $\Dctxpy$ & 46.4B & $\Denvpass$ & \textbf{53.6} & \textbf{57.8} \\
$\Denv$ + $\Dctx$ & 73.1B & $\Denvpass$ & \textbf{56.1} & \textbf{58.5} \\
\bottomrule
\end{tabular}
\caption{Ablation of data components in MT. \textbf{Top:} In a zero-shot setting (no SFT), grounding trajectories with PR data yields massive gains (+7.7\% on 72B). \textbf{Bottom:} Even when performing SFT, exposing the model to trajectories during MT improves final performance (comparing rows 3 and 4).}
\label{tab:synergy}
\end{table}

\paragraph{Trajectories require PR grounding.}
In the zero-shot setting (top section), training on environmentally-native trajectories alone yields 47.1\% (72B). However, mixing in the Python contextually-native subset $\Dctxpy$ boosts performance to 54.8\%---a significant \textbf{+7.7\%} gain. This suggests that while environmentally-native trajectories teach the model \emph{how} to interact with the environment, contextually-native data provides the necessary \emph{knowledge} and code modification diversity required to solve complex issues.

\paragraph{Mid-training on trajectories aids SFT.}
A key question in agent training is whether ``double-dipping''---training on trajectories during MT and then fine-tuning on them during SFT---provides value. Comparing the first two rows of the SFT section (Table~\ref{tab:synergy}), we observe a consistent improvement when trajectories are included in MT. For the 72B model, adding trajectories to the MT mix improves the final SFT score from 56.5\% to \textbf{57.8\%}. This indicates that mid-training allows the model to internalize the dynamics of the execution environment more deeply than SFT alone, creating a better initialization for the final alignment stage. Finally, our strongest result, \textbf{58.5\%} (72B) and \textbf{56.1\%} (32B) comes from scaling the contextually-native foundation from $\Dctxpy$ (41.9B) to the full $\Dctx$ (68.6B). This demonstrates that while mixing trajectories into MT is beneficial, the sheer scale and diversity of contextually-native supervision remain the dominant factors in model performance.

\subsection{Scalability: from raw PRs to executable tasks}
\label{sec:analysis_scalability}

Our approach is scalable along two axes: data availability and empirical performance scaling.

\begin{figure}[t]
    \centering
    \includegraphics[width=0.85\linewidth]{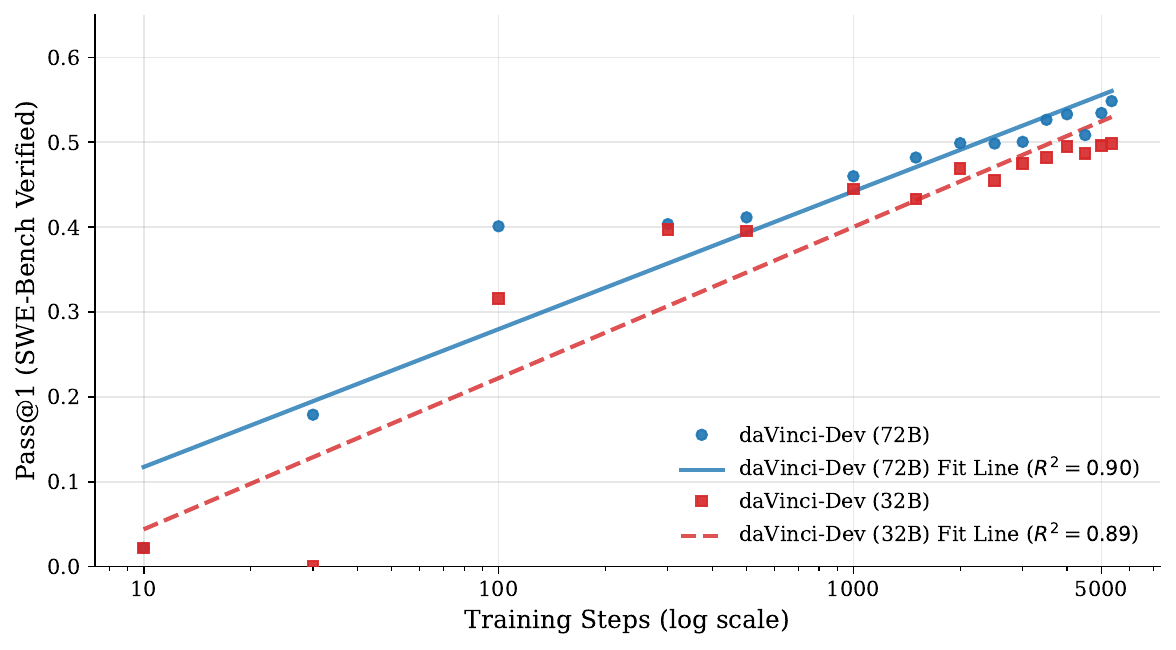}
    \caption{\textbf{Scaling Law of Agent-Native Mid-training.} Pass@1 performance on \texttt{SWE-Bench Verified} during mid-training (MT) on the $\Dctxpy$ + $\Denv$ mixture. The strong log-linear fit indicates that agentic capabilities scale predictably with training steps and data consumption, suggesting the model has not yet saturated.}
    \label{fig:scaling_curve}
\end{figure}

\paragraph{Empirical Scaling.}
Beyond the theoretical abundance of data, we verify that our model effectively converts additional compute and training steps into performance gains. Figure~\ref{fig:scaling_curve} illustrates the learning curves of both \texttt{Qwen2.5-72B} and \texttt{Qwen2.5-32B} during the MT stage on the $\Dctxpy$ + $\Denv$ mixture. We observe a robust log-linear relationship between training steps and Pass@1 performance for both model sizes ($R^2 \approx 0.90$). Specifically, the 72B model climbs to \textbf{54.9\%}, while the 32B model follows a parallel trajectory to reach \textbf{49.9\%}. Notably, we select the $\Dctxpy$ + $\Denv$ mixture as training dataset in this experiment because only when $\Denv$ is added can the models achieve zero-shot agentic capabilities (without SFT) directly from mid-training, and that we train the model using only one stage \ref{app:training_staging}. The consistent monotonic upward trends, suggests that performance has not saturated. This indicates that further scaling would yield continued improvements.

\paragraph{Scaling PR data.}
The Python-focused subset $\Dctxpy$ is built from $\sim 1.3 \times 10^{7}$ pull requests (before filtering) in $\sim 7.4\times 10^{5}$ repositories while the multi-language subset $\Dctxgen$ only utilize the $1 \times 10^4$ most starred repositories. However, our survey indicates there are $\sim 3 \times 10^{8}$ pull requests in $\sim 10^{9}$ public repositories, suggesting substantial headroom to scale the corpus by expanding language coverage and relaxing filters.

\paragraph{Scaling executable supervision.}
Recent advances in environment construction~\citep{badertdinov2025scaling} demonstrate that PR-derived data can be automatically transformed into \emph{deeper} supervision: executable tasks with Docker environments and unit tests, constructed via a fully automated pipeline. Conceptually, this is a more processed form of PR data than our raw PR corpora. Because verification is test-based and environments are executable, scaling increases not only quantity but also authenticity: each additional task comes with real environment feedback rather than synthetic traces. Importantly, \textsc{SWE-rebench} builds its public dataset from \textbf{3,468} Python repositories and reports \textbf{21,336} validated tasks, suggesting substantial headroom for scaling as repository coverage expands. This supports a long-term path where raw PR mining provides breadth, while rebench-style processing provides depth and verifiability for agent training.

\section{Limitations}
\label{sec:limitations}

\paragraph{Data privacy and attribution.}
We did not explicitly remove developer identifiers from PR text in the general subset $\Dctxgen$, which may raise privacy concerns and could lead to memorization of contributor names.

\paragraph{Evaluation sensitivity.}
Some results depend on a patched evaluation harness that fixes a small number of benchmark issues. This introduces an additional source of variance.

\paragraph{Scope.}
We focus on a single base model family and a single benchmark. Extending to other model families and more real-world agentic tasks~\citep{li2026agencybench, xu2025researcherbenchevaluatingdeepai, wu2025innovatorbench}  is left for future work.

\section{Related Work}
\label{sec:related_work}

\paragraph{Mid-training}
Recent work increasingly positions mid-training as a critical bridge between large-scale pre-training and post-training. Rather than transitioning directly from noisy, web-scale corpora to SFT or RL, mid-training introduces higher-quality, task-structured, or instruction-oriented data at later stages of training, often paired with learning-rate annealing~\citep{mo2025midtraininglargelanguagemodels,zhang2025interplaypretrainingmidtrainingrl,tu2025surveyllmmidtraining}. For example, OctoThinker~\citep{wang2025octothinkermidtrainingincentivizesreinforcement} argues that mid-training can substantially improve both the sample efficiency and the achievable performance ceiling of subsequent RL by stabilizing internal representations and encouraging reasoning-friendly behaviors.
Despite these advances, existing studies provide only limited insight into agentic mid-training data. For example, although Kimi-Dev~\citep{yang2025kimidevagentlesstrainingskill} incorporates data such as file retrieval and file editing during mid-training, these behaviors are treated in isolation and do not constitute a coherent, end-to-end agentic process. In contrast to prior work, this paper introduces Agent-Native Mid-Training, a paradigm that treats agentic behavior as a first-class training objective. We systematically design and construct mid-training data that reflects complete agentic processes and release both the construction methodology and the resulting datasets to the community.

\paragraph{Agentic training} Agentic training builds upon prior work, SFT and RL. Early agents were predominantly trained by sampling trajectories in specific environments using closed-source large models, followed by applying SFT to distill the collected data into smaller, task-specialized models~\citep{zeng-etal-2024-agenttuning,chen2024agentflandesigningdatamethods,xi2024agentgymevolvinglargelanguage,fu2025interactionintelligenceiiasynchronous}. SWE-smith~\citep{yang2025swesmithscalingdatasoftware}, BugPilot~\citep{sonwane2025bugpilotcomplexbuggeneration}, SWE-rebench~\citep{badertdinov2025swerebenchautomatedpipelinetask}, and SWE-Factory~\citep{guo2026swefactoryautomatedfactoryissue} use these ideas to create datasets in the domain of code agents. With the introduction of GRPO~\citep{shao2024deepseekmathpushinglimitsmathematical,liu2025deepseek}, recent work has increasingly focused on training agents capable of multi-step reasoning, tool usage, and explicit interaction with external environments~\citep{team2025kimi,zheng2025deepresearcherscalingdeepresearch,li2025torlscalingtoolintegratedrl}.
Despite a growing body of research on agentic post-training, systematic studies of agentic mid-training remain notably scarce. Since mid-training can gain more diverse data than the post-training stage, exploring its role and potential benefits becomes important.

\paragraph{Data synthesis}

Early approaches to synthetic data primarily focused on recombining and rewriting large-scale corpora and using reject-sampling to get the final data~\citep{yuan2023scaling}.
As the demand for data scale and coverage increased, persona-driven synthesis was introduced~\citep{ge2025scalingsyntheticdatacreation,fu2025agentrefine}, enabling a systematic expansion of the task space beyond naturally occurring data.
More recently, with the rise of agent-oriented research, synthetic data has shifted from text-level generation to the synthesis of agentic processes. Through interaction within synthetic environments~\citep{liu2025deepseek,team2025kimi,badertdinov2025swerebenchautomatedpipelinetask}, models actively generate data containing decision-making trajectories, feedback loops, and long-horizon dependencies.
In this paper, synthetic data for mid-training has become an emerging trend. Mid-training synthesis emphasizes agentic data to shape intermediate representations, serving as a critical bridge between pre-training and post-training.

\section{Conclusion}
\label{sec:conclusion}

In this work, we demonstrated that the agentic coding capabilities of large language models can be substantially enhanced through a rigorous data-centric strategy leveraging GitHub pull requests and executable interaction trajectories. By constructing a unified training recipe that combines 68.6B tokens of context-rich PR data with high-quality, verified rollouts collected in executable environments, we obtained a \texttt{daVinci-Dev-72B} with strong performance on \texttt{SWE-Bench Verified} (58.5\%), surpassing recent baselines such as Kimi-Dev.

Across experiments, the key driver is agent-native data---agent-native PR supervision (context-complete samples) plus environmentally-native trajectories (executable, test-verified rollouts).

Our analysis highlights two critical insights. First, the structural representation of PR data is paramount; keeping relevant file contents and commit edits together provides a cohesive supervision signal that mirrors the ``localize-read-edit'' loop of code agents, proving more effective than decomposing PRs into isolated subtasks. Second, not all agentic trajectories are equal. We showed that training on executable, test-verified passing trajectories yields significantly higher gains than training on static or simulated traces. The synergy between these data sources—using PRs to establish general software engineering priors and verified trajectories to specialize agentic behavior—offers a highly token-efficient path to strong performance.

Looking forward, these results suggest a scalable paradigm for future code agent development. With the vast availability of public repositories and the increasing feasibility of automated environment verification, there is substantial headroom to expand this approach to broader language ecosystems and more complex software maintenance tasks. As the field shifts from single-turn code generation to autonomous engineering, bridging the gap between static historical data and dynamic execution environments will be essential.

\section*{Acknowledgments}

We express our gratitude to Haoyang Zou, Zengzhi Wang, and Fan Zhou for their constructive feedback and stimulating discussions. We are also grateful to Liming Liu for his guidance and advice.

\bibliographystyle{acl_natbib}
\bibliography{bib}

\appendix

\section{Training Details}
\label{app:training_details}

\subsection{Dataset Components and Staging}
\label{app:training_staging}

\paragraph{PR MT Staging.}
Our $\Dctx$ (68.6B) training was conducted in two sequential stages rather than a single mix. We first trained on the general subset $\Dctxgen$ (26.7B tokens) to establish a broad software engineering baseline. We then performed mid-training (MT) on the Python subset $\Dctxpy$ (41.9B tokens) to specialize the model on agent-native, Python-centric patterns. Our $\Dctx$ + $\Denv$ (73.1B) was also conducted in two sequential stages where the first stage is the general subset $\Dctxgen$ (26.7B tokens) and the second stage is the other two datasets.

\paragraph{SFT Configuration.}
For all SFT experiments involving our $\Denvpass$ or $\Dswe$ datasets, we trained for \textbf{5 epochs}.

\subsection{Hyperparameters}
\label{app:hyperparams}

We provide the key hyperparameters used for Mid-training (MT) and Supervised Fine-tuning (SFT) below.

\paragraph{MT Hyperparameters.}
We use a global batch size of 1024 samples and a peak learning rate of $8 \times 10^{-5}$. The learning rate schedule utilizes a warmup ratio of 0.05 (5\% of total training steps), followed by cosine decay until all samples are consumed once (1 epoch). No loss mask is applied during MT.

\paragraph{SFT Hyperparameters.}
We use a global batch size of 128 samples and a peak learning rate of $1 \times 10^{-5}$. The learning rate schedule utilizes a warmup ratio of 0.10 (10\% of total training steps), followed by cosine decay until all samples are consumed once per epoch. A standard loss mask is applied to user and tool tokens during SFT.

\section{LLM prompts used for PR rendering}
\label{app:llm_prompts_pr_rendering}

During context enrichment (Section~\ref{sec:contextually_native}), we optionally call an LLM to (i) generate a concise pull-request summary and (ii) normalize/optimize commit messages for readability. We use \texttt{Qwen3-235B-A22b-Instruct-2507} with fixed output budgets (512 tokens for PR summaries; 256 tokens for commit-message refinement).

\begin{figure}[htbp]
\begin{tcolorbox}[
colback=gray!5,
colframe=gray!75,
left=2mm, right=2mm,
title=\textbf{PR Summary Prompt}]
\small
\begin{verbatim}
Summarize this pull request in 1-4 clear sentences:

Repository: {{.RepoName}}
Description: {{.RepoDesc}}

PR Title: {{.Title}}
PR Description:
{{.Body}}

{{if .Issue}}Related Issue: {{.Issue.Title}}
{{.Issue.Body}}

{{end}}Changed Files:
{{range .ChangedPyFiles}}- {{.}}
{{end}}

Commits:
{{range .Commits}}
## Message: {{.Message}}

Changes:
{{range .Diffs}}
File: {{.Path}}
{{.Patch}}
{{end}}
{{end}}

Please provide a clear and concise summary (1-4 sentences) of this Pull Request,
focusing on:
1. What problem does it solve or what feature does it add?
2. What are the key changes made?
3. Any important implementation details?

Summary:
\end{verbatim}
\end{tcolorbox}

\begin{tcolorbox}[
colback=gray!5,
colframe=gray!75,
left=2mm, right=2mm,
title=\textbf{Commit Message Refinement Prompt}]
\small
\begin{verbatim}
Optimize this commit message for clarity and educational value while keeping it
concise.

PR Context Summary: {{.Summary}}

Original commit message:
{{.Commit.Message}}

Diff Context:
{{range .Commit.Diffs}}File: {{.Path}}
{{truncatePatch .Patch 2000}}

{{end}}
Provide an optimized version that:
1. The subject is clear and descriptive
2. If the commit is trivial and the changes are minimal, don't add the footer
3. Otherwise, keep the footer in one sentence

Refined commit message:
\end{verbatim}
\end{tcolorbox}
\caption{Prompts used for optional LLM-based PR summary generation and commit-message refinement during PR rendering.}
\label{fig:llm_prompts_pr_rendering}
\end{figure}

\section{Dataset Formats}
\label{app:pr_data_template}

We include the templates for two types of data in the contextually-native dataset: (i) the \textbf{General PR} format, and (ii) the \textbf{Python PR} agentic format.

The General PR format uses XML-like tags similar to \texttt{The Stack v2}~\citep{lozhkov2024starcoder2stackv2} and includes rich interaction history (comments and reviews). Events related to a pull requests are concatenated in chronological order. Different from \texttt{The Stack v2}, we always include relevant file content and grouped review comments threads. This corpus is sourced from top-starred repositories without the 1--5 Python-file constraint.

\begin{figure}[htbp]
\begin{tcolorbox}[
colback=gray!5,
colframe=gray!75,
left=2mm, right=2mm,
title=\textbf{General PR Example}]
\small
\textbf{\# Repository Context}

Name: parcel-bundler/parcel\\
Description: The zero configuration build tool for the web.

\textbf{\# Relevant Files Context}

\textbf{\#\# packages/core/parcel-bundler/src/cli.js}

\begin{verbatim}
...
  if (command.name() === 'serve' && command.target === 'browser') {
    const server = await bundler.serve(
      command.port || 1234,
      command.https,
      command.host
    );
...
\end{verbatim}

\textbf{Response:}

\textless pr\textgreater Title: use env port\\
lizzzp1: Adds \texttt{process.env.PORT} as a default port option...

\textless pr\_comment\textgreater mischnic: I think it should rather be...\\
\textless pr\_review\textgreater devongovett: Looks good to me.\\
\textless pr\_review\_state\textgreater approved

\textless pr\_commit\textgreater Liz P: use env port\\
\textless commit\_file\textgreater packages/core/parcel-bundler/src/cli.js\\
\textless patch\textgreater
\begin{verbatim}
@@ -219,7 +219,7 @@ async function bundle(main, command) {
   if (command.name() === 'serve' && command.target === 'browser') {
     const server = await bundler.serve(
-      command.port || 1234,
+      process.env.PORT || 1234,
       command.https,
       command.host
    );
\end{verbatim}
\textless /patch\textgreater

\textless pr\textgreater devongovett\\
\textless pr\_status\textgreater closed\\
\textless pr\_is\_merged\textgreater True

\end{tcolorbox}
\caption{Example of the general PR format.}
\label{fig:format_general_pr}
\end{figure}

The Python PR format uses a Markdown structure and represents edits in a \emph{search-and-replace} action space. It includes an LLM-generated PR summary after presenting all related files (simulating a overall planning and reasoning phase in an agentic workflow) and enhanced commit messages (simulating textual reasoning before action). Edits are rewritten from git diff format to search-replace format used in many agentic scaffolds. The \textbf{\# Issue} section is omitted if no linked issue is found.

\begin{figure}[htbp]
\begin{tcolorbox}[
colback=gray!5,
colframe=gray!75,
left=2mm, right=2mm,
title=\textbf{Python PR Example}]
\small
\textbf{\# Repository Context}

Name: Pylons/waitress\\
Description: Waitress - A WSGI server for Python 3

\textbf{\# Issue}

\textbf{\#\# \textbackslash xa0 and \textbackslash x85 are stripped from header values}\\
Given that these bytes are allowed in header values (due to \texttt{obs-text}), they shouldn't be stripped during header-field OWS stripping...

\textbf{\# Pull Request}

\textbf{\#\# Bugfix: Don't strip whitespace from values before inserting into environ}\\
This fixes a small bug where the value of the header would get stripped when inserted into the environ so it no longer matched. Closes \#432

\textbf{\# Relevant Files Found}

\textbf{\#\# src/waitress/task.py}

\begin{verbatim}
```
...
        for key, value in dict(request.headers).items():
            value = value.strip()
            mykey = rename_headers.get(key, None)
...
```
\end{verbatim}

\textbf{\# Edits}

This pull request removes the erroneous \texttt{.strip()} call on header values in the WSGI environ construction. The HTTP specification allows certain non-ASCII bytes (\texttt{\textbackslash xa0}, \texttt{\textbackslash x85}) in header values via \texttt{obs-text}, and these should not be stripped.

Remove the strip() call from header value processing in get\_environment()

Edit: src/waitress/task.py

Search:
\begin{verbatim}
```
        for key, value in dict(request.headers).items():
            value = value.strip()
            mykey = rename_headers.get(key, None)
            if mykey is None:
                mykey = "HTTP_" + key
```
\end{verbatim}

Replace:
\begin{verbatim}
```
        for key, value in dict(request.headers).items():
            mykey = rename_headers.get(key, None)
            if mykey is None:
                mykey = "HTTP_" + key
```
\end{verbatim}

\end{tcolorbox}
\caption{Example of the Python PR agentic format.}
\label{fig:format_python_pr}
\end{figure}

\section{Benchmark decontamination}
\label{app:benchmark_decontamination}

In our training dataset we take measures to remove samples related to the \texttt{SWE-Bench Verified} benchmark as detailed in Section~\ref{sec:contextually_native}. For the HumanEval and EvalPlus benchmarks, we follow the decontamination procedure of \texttt{XCoder}~\citep{wang2024codellmsperformempowering}. Concretely, for each benchmark instance we form the reference text by concatenating the prompt and canonical solution, tokenize it, and compute the set of unique $n$-grams ($n=13$). We then scan the tokenized training corpus and, for every training sample, compute its set of unique 13-grams and the overlap with each benchmark instance. Similarity is measured as a leakage ratio:
\[
\mathrm{leakage\_ratio}(e, x) \;=\; \frac{\left|G_e \cap G_x\right|}{\left|G_e\right|},
\]
where $G_e$ is the set of unique 13-grams in the benchmark instance and $G_x$ is the set of unique 13-grams in a training sample. For each benchmark instance, we take the maximum leakage ratio over all training samples as its contamination score.

We manually selected the contamination threshold as $\tau=0.10$ based on case studies of high-overlap matches. Using this criterion, we identified 24 contaminated HumanEval instances, which were removed from evaluation.

\end{document}